\documentclass[12pt]{article}

\usepackage{amssymb}
\usepackage{amsmath}
\usepackage{amsfonts}
\usepackage{epsfig}
\usepackage{anysize}
\marginsize{2cm}{2cm}{2.5cm}{3cm}
\date{}

\usepackage{hyperref}

\def\captionFigureI{%
  \caption{\label{Fig1}\footnotesize
  \textbf{Two examples of ranked data.}
 (a) Data for the energy released by earthquakes in California
\cite{quakes1}. (b) Same earthquake data ranked according to number of
occurrences of earthquakes of similar magnitude showing behavior
compatible with the Guttenberg-Richter law. (c) Data for the areas
burnt in forest fires in the U.S.A. \cite{fires1}. (d) Same
forest fires data ranked according to the number of occurrences of
similar burnt areas. See text for description.}
}

\def\captionFigureII{%
  \caption{\label{Fig2}\footnotesize
    \textbf{Theoretical fitting of ranked data.}
Same two examples in Fig 1 of ranked data on earthquakes and
forest fires fitted with the expressions in Eqs. (\ref{size-rank1})
and (\ref{frequency-rank1a}). (a) Size-rank distribution $N(k)$ for
earthquakes. (b) Frequency-rank distribution for earthquakes (with
$F(k^{\prime })=\mathcal{N}f(k^{\prime })$). (c) Size-rank distribution
$N(k)$ for forest fires. (d) Frequency-rank distribution for forest
fires (with $F(k^{\prime })=\mathcal{N}f(k^{\prime })$). As can be
seen, the values of $\alpha$ needed for fitting are close to
$\alpha=2$ that corresponds to the classical Zipf law exponent. See
text for description.}
}

\def\captionFigureIII{%
  \caption{\label{Fig3}\footnotesize
    \textbf{Iterated map at tangency.}
The map parameters are
 $z=\alpha=2$,
 the curvature is $u=0.0125$, and the trajectory $x_{t}, t=0,1,2,\ldots
t$, is initiated at $x_{0}$ as given by
Eq (\ref{trajectory}). Also shown is the area $A_{t}$ (shaded) as
given by Eq (\ref{area1}). The map properties translate into the
equivalent description of the rank distributions $N(k)$ and
$F(k^{\prime})$ via the identifications $t=k$, $u=\mathcal{N}^{-1}$,
$x_{t}=-N(k)=k^{\prime}$, $x_{0}=-N_{\max }$ and
$A_{t}=F(k^{\prime}/\mathcal{N})$. See text for description.}
}

\def\captionFigureIV{%
  \caption{\label{Fig4}\footnotesize
    \textbf{Rank distributions for Benford law.}
Rank distributions for Benford law together with numerical
data that follows this law (shown in the inset). (a) Frequency-rank distribution
$F(k^{\prime })$. (b) Size-rank distribution $N(k)$. They are obtained from the
general formalism with $\alpha=1$. Data taken from Table I in the original
Benford's article Ref. \cite{benford2}. See text for description.}
}

\begin{document}
\renewcommand{\figurename}{\textbf{\footnotesize Figure}}
\renewcommand{\thefigure}{\mbox{\textbf{\footnotesize\arabic{figure}}}}

\title{\LARGE\bf Rank Distributions: Frequency vs.\ Magnitude}

\author{Carlos Velarde\textsuperscript{1}, Alberto Robledo\textsuperscript{2}\\
\footnotesize 1. Instituto de Investigaciones en Matem\'aticas Aplicadas y en Sistemas,\\%
\footnotesize    Universidad Nacional Aut\'onoma de M\'exico\\%
                 %Apartado Postal 70-221, M\'exico 04510 Distrito Federal, M\'exico\\
\footnotesize 2. Instituto de F\'{i}sica y Centro de Ciencias de la Complejidad,\\%
\footnotesize    Universidad Nacional Aut\'onoma de M\'exico,\\%
\footnotesize    Apartado Postal 20-364, CDMX 01000, M\'exico
}

\maketitle

\abstract{%
We examine the relationship between two different types of ranked data,
frequencies and magnitudes. We consider data that can be sorted out either
way, through numbers of occurrences or size of the measures, as it is the
case, say, of moon craters, earthquakes, billionaires, etc. We indicate that
these two types of distributions are functional inverses of each other, and
specify this link, first in terms of the assumed parent probability distribution that
generates the data samples, and then in terms of an analog (deterministic)
nonlinear iterated map that reproduces them. For the particular case of hyperbolic
decay with rank the distributions are identical, that is, the classical Zipf plot,
a pure power law. But their difference is largest when one displays logarithmic
decay and its counterpart  shows the inverse exponential decay, as it is the
case of Benford law, or viceversa. For all intermediate decay rates generic
differences appear not only between the power-law exponents for the midway
rank decline but also for small and large rank. We extend the theoretical
framework to include thermodynamic and statistical-mechanical concepts,
such as entropies and configuration.
}

%\keywords{Physics, Statistical physics, Nonlinear physics, Nonlinear dynamical systems}

\normalsize

\section{Introduction}
Ranking data that originates from apparently disconnected subjects in many fields
---astrophysical, geophysical, ecological, biological, technological, financial, urban,
social, etc.--- has revealed universal patterns \cite{ranking1, newman1} and opened
intriguing questions about their origin. The empirical law of Zipf \cite{zipf1,zipf2} for the
numbers of occurrence (frequencies if normalized) of words in texts has played a central
role in the development of this widespread research topic of multidisciplinary complex
systems. Zipf's law has been found to be (approximately) followed by many sets of
ranked data outside linguistics, that record the number of occurrences \cite{frequencyrank1}
of other types of items. But also, and this is an important distinction we address here, for
the magnitudes or sizes of many measurable objects or entities, such as firmament voids,
lengths of rivers, city populations, etc. \cite{sizerank1}. 
 
Here we analyze the conceptual, and also quantitative, difference between frequency
and size ranked data. To this purpose we make use of a straightforward stochastic
procedure \cite{pietronero1,robledo1,robledo2} to reproduce ranked data from an assumed
parent distribution that governs sets of values of random variables that constitute samples.
Examination of the expressions for the two types of rank functions indicate that they are functional
inverses of each other. See also \cite{sizerank1,quantile1,egghe1}. In particular, we focus in the case where the
parent distribution $P(N)$, where $N$ is a magnitude random variable, has the power-law
form $P(N)\sim N^{-\alpha}$, $\ 1\leq \alpha <\infty$. We find that in the limit $\alpha=1$ the
size-rank distribution $N(k)$, where $k$ is the rank, decays exponentially as $k$ grows,
while the frequency-rank distribution $F(k')$ decays logarithmically as $k'$, the corresponding
rank variable, increases. On the contrary, in the limit $\alpha\rightarrow\infty$ $N(k)$ decays
logarithmically while $F(k')$ does so exponentially. The intermediate case $\alpha=2$ is the
special exponent value when both $N(k)$ and $F(k')$ decay as a power law with exponent $-1$,
the classical Zipf's power law value. To complement our description we replicate the procedure by
considering instead a starting parent distribution $Q(F)\sim F^{-\beta}$, $\ 1\leq \beta <\infty$ where
$F$ is a frequency random variable, and obtain an equivalent account with $1-\beta=1/(1-\alpha)$.

We have recently \cite{robledo1,robledo2,pnas1} shown that the above-referred stochastic approach
to size-rank distributions can be exactly represented by deterministic nonlinear one-dimensional
iterated maps close to tangency \cite{schuster1}. Here we extend this strict analogy to determine
frequency-size distributions within this nonlinear dynamical language. These distributions are
given by areas below map trajectories. To explore  the duality between size-rank $N(k)$ and
frequency-rank $F(k')$ distributions, we look at specific sets of real data that can be sorted out in
both ways, magnitudes or numbers of occurrences, such as the cases of earthquakes \cite{quakes1}
and forest fires \cite{fires1} (see Fig \ref{Fig1}), and we find agreement with the theoretical approach.
We also comment on how Benford's law \cite{benford1,benford2} for the frequency of digits corresponds
in our scheme to the case $\alpha=1$.

%FigureI %<<<<<<<<<<<<
\begin{figure*}[!h]\scriptsize
  \vspace*{.05in}
  \centering
  \includegraphics[width=0.75\textwidth]{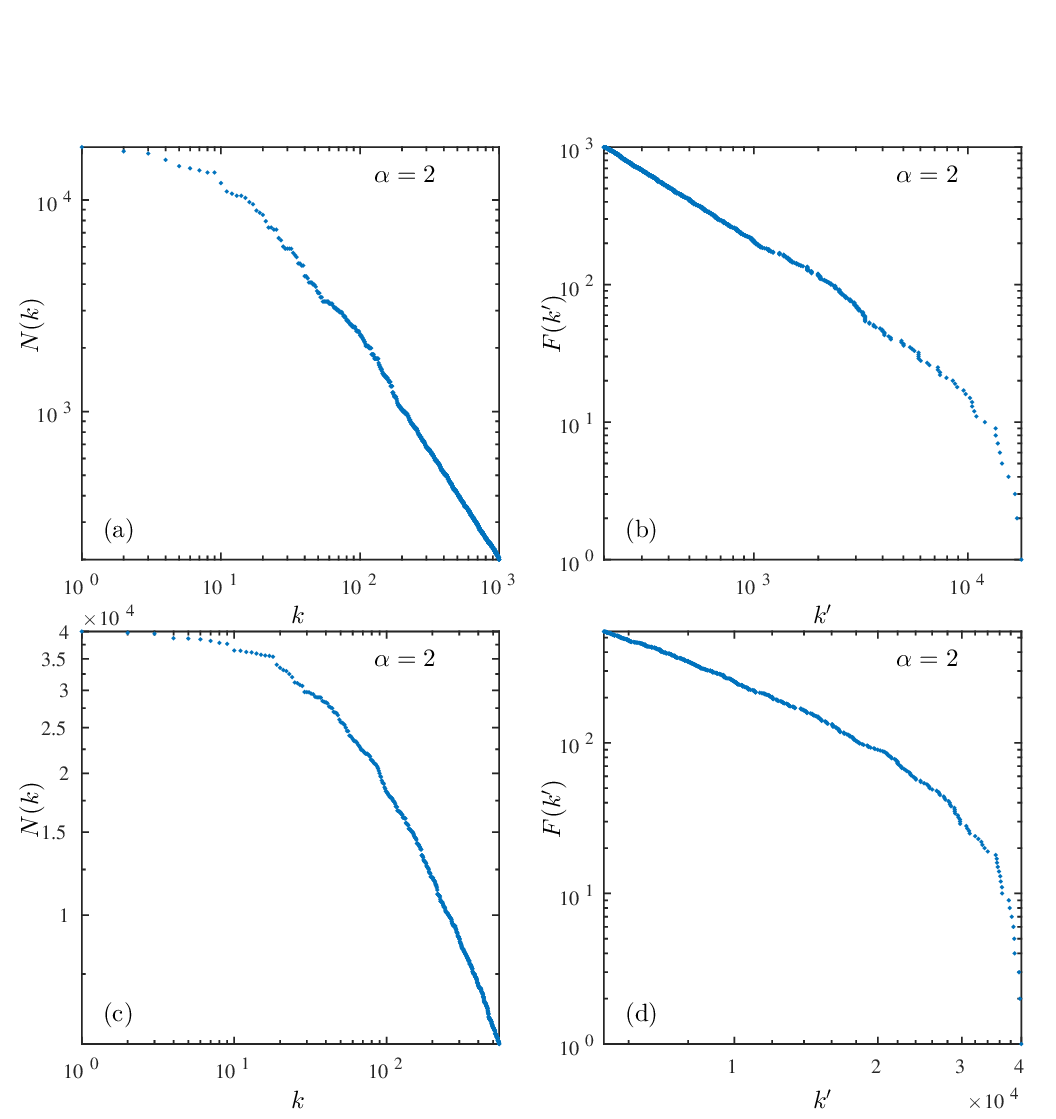}
  \parbox{0.9\textwidth}{\captionFigureI}
  %\label{Fig1}     %FigureI(Data)
\end{figure*}

Finally, we extend our statistical-mechanical interpretation with generalized entropies of rank distributions
\cite{pnas1,heliyon1} to include the role of $F(k')$.  

\section{Rank distributions from a size parent distribution}

The basic ingredient in the stochastic method \cite{pietronero1, robledo1, robledo2}
for rank distributions is the probability distribution $P(N)$ of the magnitude or size data
$N$ under consideration. The scheme is phenomenological since the form of $P(N)$ is
assumed, and so, the first common choices are: gaussian, exponential, or
power law expressions. For the latter case we write%
\begin{equation}
P(N)\sim N^{-\alpha },\ 1\leq \alpha <\infty .  \label{basic}
\end{equation}%
Sets of data $N$ can be generated from Eq (\ref{basic}) and subsequently
examined if they match, statistically, real ranked data sets. Each data set formed
by a total of $\mathcal{N}$ entries, expressed with given suitable precision,
can be ranked according to their sizes $N$ or the numbers of times $F$ with
which their items appear. We shall consider that $N$ takes positive values within
an interval $N_{\min }\leq N\leq N_{\max }$, where we allow as limiting values
$N_{\min }=0$\ and/or $N_{\max }\rightarrow\infty $. To obtain the number of
occurrences $F$ for real numbers $N$ recorded with a given precision it may
be necessary to introduce a partition and count incidences within intervals. 

The entries in the sample set $\mathcal{N}$ can be sorted out starting
with the largest, $N_{\max }$,\ and continuing with decreasing magnitudes 
down to $N_{\min }$. And then labeled with the rank variable variable $k$, with
$k=0$ for $N_{\max }$ and $k=k_{\max }$ for $N_{\min }$. We call the function
$N(k)$ the size-rank distribution. The rank $k$ can be an integer $%
k=0,1,2,3,\ldots,k_{\max }$ (often, elsewhere, the 1st value is $k=1$) and it can be
generalized to be a real number. The set $\mathcal{N}$ can also be ordered
in terms of the frequency with which they appear, that is, the number of
occurrences $F$ having size equal or greater than $N$, or equivalently the
rate $f$, $0\leq f\leq 1$, of occurrences having size equal or greater than $%
N$. For this second sorting the occurrences are labeled with a rank
variable $k^{\prime }$, with $k^{\prime }=0$ for the most frequent and $%
k^{\prime }=k_{\max }^{\prime }$ for the least frequent. We call $%
F(k^{\prime })$ the frequency-rank distribution. Similarly, the rank $%
k^{\prime }$ can be an integer $k'=0,1,2,3,\ldots,k^{\prime }_{\max }$ (often the 1st
value is $k^{\prime }=1$) but it can be generalized to be a real number.
The normalized frequency-rank distribution is $f(k^{\prime })= F(k^{\prime })/\mathcal{N}$.
The main task is to determine $N(k)$ and $F(k^{\prime })$ from $P(N)$.

We now introduce the complementary cumulative distribution of $P(N)$,%
\begin{equation}
\Pi (N,N_{\max })=\int\limits_{N}^{N_{\max }}P(N^{\prime })dN^{\prime },
\label{cumulative}
\end{equation}%
where the normalization of $P(N)$ implies $\Pi (N_{\min },N_{\max })=1$. The
parent distribution $P(N)$ can be recuperated from $\Pi (N,N_{\max })$ via%
\begin{equation}
P(N)=-\frac{\partial }{\partial N}\Pi (N,N_{\max }).  \label{basic2}
\end{equation}%
In the theoretical approach the evaluation of $\Pi (N,N_{\max })$ is the means by
which the values $N$ generated by $P(N)$ are sorted out and leads to the rank distributions.

The cumulative distribution $\Pi (N,N_{\max })$ increases
monotonically as $N$ decreases, taking values from $\Pi (N_{\max },N_{\max })=0$
to $\Pi (N_{\min },N_{\max })=1$. This distribution $\Pi (N(k),N_{\max })$, where
we have now indicated the rank $k$ occupied by the variable magnitude $N$,
is identified with $k/\mathcal{N}$, that is%
\begin{equation}
\frac{k}{\mathcal{N}}\equiv\Pi (N(k),N_{\max }).  \label{rank}
\end{equation}%
\ The size-rank distribution $N(k)$ is obtained by solving%
\begin{equation}
\frac{k}{\mathcal{N}}=\int\limits_{N(k)}^{N_{\max }}P(N^{\prime })dN^{\prime
},  \label{cumulative2}
\end{equation}%
for $N(k)$. Normalization of $P(N)$ indicates that $k_{\max }=\mathcal{N}$.
If $k$ is to be an integer the possible lower limits in the integral in Eq (%
\ref{cumulative2}), $N(1)$, $N(2)$, $\ldots$, $N(k_{\max })$ are such that the
integral takes values $1/\mathcal{N}$, $2/\mathcal{N}$, $\ldots$, $k_{\max }/%
\mathcal{N}$.

On the other hand, the fraction $k/\mathcal{N}$ can also be seen as the rate
or scaled frequency with which the sizes equal or greater than $N$ occur,
small for small $k\simeq 0$ and large for $k\simeq k_{\max }$. Therefore we
identify the normalized frequency-rank distribution $f(k^{\prime })$ as%
\begin{equation}
f(k^{\prime })\equiv\Pi (N,N_{\max }),  \label{cumulative3}
\end{equation}%
where $k^{\prime }\equiv N$. If $k^{\prime }$ is to be an integer the values of $N$
to be used in $P(N)$ are integers. In practice, the non-normalized frequency-size
distribution $F(k^{\prime })\equiv\mathcal{N}f(k^{\prime })$ is often used as it is
constructed directly from the numbers of occurrences in data samples. From the
above definitions $k^{\prime }\equiv N$, and $F(k^{\prime })\equiv\mathcal{N}f(k^{\prime })$,
together with Eqs. (\ref{rank}) and (\ref{cumulative3}), it is clear that the
rank distributions $N(k)$ and $F(k^{\prime })$ are functional inverses of each
other. That is, $k=F(N)$ or $N=F^{-1}(k)$. The inverse of a cumulative distribution
is referred to as the quantile function \cite{sizerank1,quantile1}. We refer to $N(k)$ as the
size-rank distribution even though technically it is not a probability distribution, as
$P(N)$ and $f(k^{\prime })$ are. 

\section{Rank distributions from a power-law parent distribution}

We look now at the specific expressions that come out of the general equations
in the previous Section when $P(N)$ is given by Eq (\ref%
{basic}). We have%
\begin{eqnarray}
\Pi (N(k),N_{\max }) &=&\int\limits_{N(k)}^{N_{\max }}N^{-\alpha }dN 
\nonumber \\
&=&\frac{1}{1-\alpha }\left[ N_{\max }^{1-\alpha }-N(k)^{1-\alpha }\right] ,
\label{cumulative4}
\end{eqnarray}%
or, in \ terms of the $q$-deformed logarithmic function $\ln_{q}(x)\equiv
(1-q)^{-1}[x^{1-q}-1]$ with $q$ a real number,%
\begin{equation}
\ln _{\alpha }N(k)=\ln _{\alpha }N_{\max }-\mathcal{N}^{-1}k.
\label{logalpha1}
\end{equation}%
The size-rank distribution $N(k)$ is explicitly obtained from the above with
use of the inverse of $\ln _{q}(x)$, the $q$-deformed exponential function $%
\exp _{q}(x)\equiv \left[ 1+(1-q)x\right] ^{1/(1-q)}$, this is%
\begin{equation}
N(k)=N_{\max }\exp _{\alpha }[-N_{\max }^{\alpha -1}\mathcal{N}^{-1}k].
\label{size-rank1}
\end{equation}%
While the frequency-rank distribution $f(k^{\prime })$ is given by%
\begin{eqnarray}
f(k^{\prime }) &=&\ln _{\alpha }N_{\max }-\ln _{\alpha }k^{\prime } 
\nonumber \\
&=&1+\ln _{\alpha }N_{\min }-\ln _{\alpha }k^{\prime }.
\label{frequency-rank1a}
\end{eqnarray}

In Fig \ref{Fig2} we show the agreement of Eqs (\ref{size-rank1}) and (\ref{frequency-rank1a})
with the data on earthquakes and forest fires already shown in Fig \ref{Fig1}. 
% new text
Our method for fitting the data to Eqs (\ref{size-rank1}) and (\ref{frequency-rank1a}) is heuristic.
We first select a data point to define $N_{\max}$. We then approximate with a straight line segment a section of
the data that appears lined when displayed in logarithmic scales (involving a choice of its two extremes) via
minimum squares. This gives us, with the use of Eq. (\ref{logalpha1}), a set of two equations from which we
determine numerically preliminary values for $\alpha$ and $\mathcal{N}$ (notice that Eq (\ref{cumulative4})
has no normalization constant). We iterate this procedure to improve fitting (mostly only $\mathcal{N}$ changes
its value appreciably). Once the parameters in Eq (\ref{size-rank1}) are determined $F(k^{\prime})$ follows from
Eq (\ref{frequency-rank1a}). 

%FigureII %<<<<<<<<<<<<
\begin{figure*}[!h]\scriptsize
  \vspace*{.05in}
  \centering
  \includegraphics[width=0.75\textwidth]{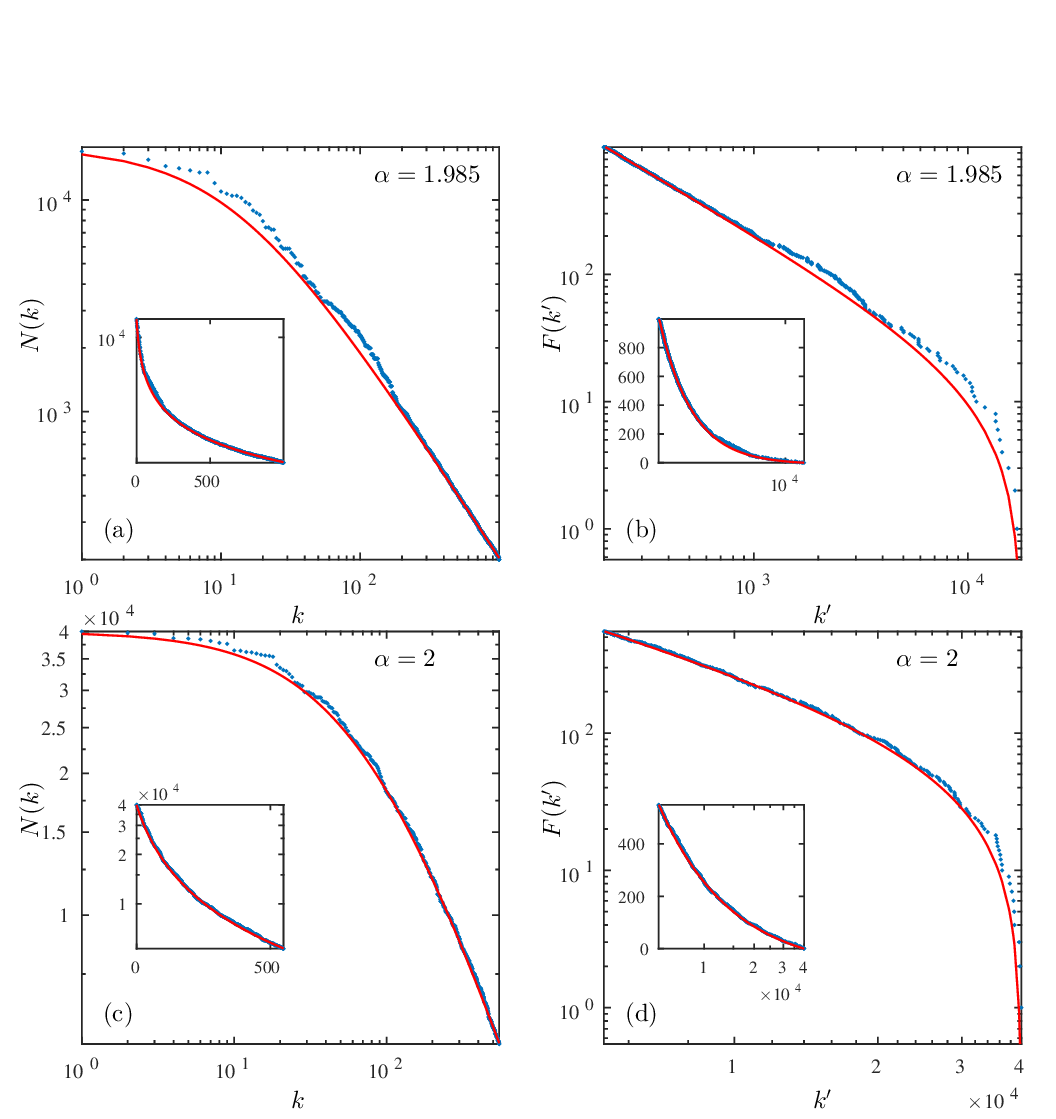}
  \parbox{0.9\textwidth}{\captionFigureII}
  %\label{Fig2}     %FigureII(Maps)
\end{figure*}

When $\alpha =1$ Eq (\ref{size-rank1}) acquires the ordinary exponential
form%
\begin{equation}
N(k)=N_{\max }\exp (-\mathcal{N}^{-1}k),  \label{size-rank2}
\end{equation}%
while Eq (\ref{frequency-rank1a}) becomes an ordinary logarithmic function, 
\begin{eqnarray}
   f(k^{\prime }) &=& \ln \ (N_{\max }/k^{\prime })  \nonumber \\
                  &=& 1-\ln \ (k^{\prime }/N_{\min }).  \label{frequency-rank2}
\end{eqnarray}%
We take the limit $\alpha \rightarrow \infty $ to signify that $%
P(N)=N_{0} \exp (- N_{0} N)$, and we choose $N_{0}=1$. We find%
\begin{equation}
N(k)=-\ln \left[ \exp (-N_{\max })+\mathcal{N}^{-1}k\right] ,
\label{size-rank3}
\end{equation}%
and%
\begin{eqnarray}
f(k^{\prime }) &=&\exp (-k^{\prime })-\exp (-N_{\max }) \\
&=&\exp (-k^{\prime })-\exp (-N_{\min })+1.  \label{frequency-rank3}
\end{eqnarray}

In the limit $N_{\max }\rightarrow \infty $ Eq (\ref{size-rank1}) becomes
the power law $N(k)\sim k^{1/(1-\alpha )}$ that when $\alpha =2$ gives the
simple hyperbolic form $N(k)\sim k^{-1}$. Whereas Eq (\ref{frequency-rank1a})
in the same limit becomes the power law $f(k^{\prime })$ $\sim k^{\prime (1-\alpha )}$
that when $\alpha =2$ gives, coincidentally, the same hyperbolic form $f(k^{\prime })$ $\sim
k^{\prime -1}$. For many sets of frequency-rank real data $\alpha \simeq 2$
and the standard Zipf law is $\alpha =2$, whereas the same feature for real
size-rank data has led to refer (concurrently) to the observation of Zipf's
law in relation to $N(k)$. In contrast, when $\alpha \rightarrow \infty $,
in the limit $N_{\max }\rightarrow \infty $ the rank distributions become $%
N(k)=\ln \ (\mathcal{N}/k)$ and $f(k^{\prime })=\exp (-k^{\prime })$, $N(k)$
decays very fast as $k$ increases since the argument in the logarithmic
function lies in the interval $0<k/\mathcal{N<}1$, while $f(k^{\prime })$
decays exponentially as $k^{\prime }$ increases. This can be compared with
the case $\alpha =1$, but $N_{\max }$ finite, when $N(k)$ decays
exponentially as $k$ increases while $f(k^{\prime })$ decays very fast as 
$k^{\prime }$ increases since again the argument in the logarithmic function
lies in the interval $0<k^{\prime }/N_{\max }\mathcal{<}1$.

A note on normalization. The choice of $P(N)$ given by Eq (\ref{basic}) is
not compatible with finite data sets ($\mathcal{N<\infty }$), these should
be represented by a different expression for $P(N)$, at least one that
differs from Eq (\ref{basic}) for some values of $N$, specially small $N$.
Normalization of Eq (\ref{basic}) obeys $k_{\max }=\mathcal{N}$, with both $%
k_{\max }\rightarrow \infty $ and $\mathcal{N\rightarrow \infty }$,
 while $N_{\min }\rightarrow 0$.

\section{Rank distributions from a frequency parent distribution}

To show a duality feature of the approach to rank distributions we now
consider the derivation of these distributions from a different parent
distribution. This distribution, $Q(F)$, generates values of the numbers
of occurrences $F$ to form data sets. As before we introduce a complementary
cumulative distribution
\begin{equation}
X (F,F_{\max })=\int\limits_{F}^{F_{\max }}Q(F^{\prime })dF^{\prime },
\label{cumulative?}
\end{equation}%
where the normalization of $Q(F)$ implies $X (F_{\min },F_{\max })=1$. We denote
by $\mathcal{F}$ the total number of elements in the occurrences sample set.

Proceeding as before we indicate the rank $k^{\prime }$ occupied by the number of
occurrences $F$ in the distribution $X (F(k^{\prime }),F_{\max })$ and identify this
as $k^{\prime }/\mathcal{F}$. That is
\begin{equation}
\frac{k^{\prime }}{\mathcal{F}}\equiv X (F(k^{\prime }),F_{\max }).  \label{rank2}
\end{equation}%
When we  assume the power law expression
\begin{equation}
Q(F)\sim F^{-\beta },\ 1\leq \beta <\infty ,  \label{basic2}
\end{equation}%
we obtain
\begin{equation}
X(F(k^{\prime }),F_{\max }) = \frac{1}{1-\beta }\left[ F_{\max }^{1-\beta }-F(k^{\prime })^{1-\beta }\right] ,
\label{cumulative5}
\end{equation}%
or, in terms of the $q$-deformed logarithmic function,
\begin{equation}
\ln _{\beta }F(k^{\prime })=\ln _{\beta }F_{\max }-\mathcal{F}^{-1}k^{\prime }.
\label{logbeta1}
\end{equation}%
The frequency-rank distribution $F(k^{\prime })$ is explicitly obtained from the above with
with use of the $q$-deformed exponential function, this is%
\begin{equation}
F(k^{\prime })=F_{\max }\exp _{\beta }[-F_{\max }^{\beta -1}\mathcal{F}^{-1}k^{\prime }].
\label{frequency-rank1b}
\end{equation}%
While the size-rank distribution $N(k)$, following arguments parallel to those given
before for $F(k^{\prime })$,  is given by
\begin{equation}
N(k)\equiv \mathcal{F} X (k,F_{\max }),  \label{cumulative??}
\end{equation}%
where $k\equiv F$. Explicitly,%
\begin{eqnarray}
N(k) &=&\ln _{\beta }F_{\max }-\ln _{\beta }k 
\nonumber \\
&=&1+\ln _{\beta }F_{\min }-\ln _{\beta }k.
\label{frequency-rank?}
\end{eqnarray}
Again, it is clear that the rank distributions $F(k^{\prime })$ and $N(k)$ are
functional inverses of each other. That is, $k^{\prime }=N(F)$ or $F=N^{-1}(k^{\prime })$.

The exponent $\alpha$ in the previous two sections and the exponent $\beta$ in
this section are related via
\begin{equation}
1-\alpha=\frac{1}{1-\beta },
\end{equation}
and coincide in value when $\alpha=\beta=2$, and both distributions acquire the simple
hyperbolic functions $F(k^{\prime })\sim {k^{\prime }}^{-1}$ and $N(k)\sim k^{-1}$ when
in addition $F_{\max }\rightarrow \infty $ and $N_{\max }\rightarrow \infty $, that is,
the classical Zipf case.

\section{Rank distributions from a nonlinear map at tangency}

We have shown recently \cite{robledo1,robledo2,pnas1} that there is an exact analogy
between the expressions for the rank distributions as presented above for $N(k)$ and
those for the trajectories associated with the tangent bifurcation in one-dimensional
nonlinear iterated maps. A map $g(x)$ at the tangent bifurcation is written locally
as $x^{\prime}=g(x)=x-u\left\vert x\right\vert ^{z}+\cdots$,$\;x\leq 0$ \cite{schuster1},$\;z>1$,
and trajectories initiated at $x_{0}\lesssim 0$ are obtained via repeated
iterations of $g(x)$, i.e. 
\begin{equation}
x_{\tau +1}=x_{\tau }-u\left\vert x_{\tau }\right\vert ^{z},\tau =0,1,\ldots
\label{tangent}
\end{equation}%
These trajectories move monotonically towards the point of tangency at $x=0$%
. If we make the replacement, valid for large time $\tau $, of the
difference $x_{\tau +1}-x_{\tau }$ by $dx_{\tau }/d\tau $ in Eq (\ref%
{tangent}) (written as $-u\left\vert x_{\tau }\right\vert ^{z}=x_{\tau
+1}-x_{\tau }$) we obtain the differential form $ud\tau =-\left\vert x_{\tau
}\right\vert ^{-z}dx_{\tau }$, and integration of both sides of it yields%
\begin{eqnarray}
ut &=&\int_{x_{0}}^{x_{t}}\frac{dx_{\tau }}{-\left\vert x_{\tau }\right\vert
^{z}}  \nonumber \\
&=&\frac{1}{1-z}\left[-\left\vert x_{t}\right\vert ^{1-z}+\left\vert
x_{0}\right\vert ^{1-z}\right] ,  \label{tangent2}
\end{eqnarray}%
or%
\begin{equation}
\ln _{z}\left\vert x_{t}\right\vert =\ln _{z}\left\vert x_{0}\right\vert
-ut.  \label{logz1}
\end{equation}%
The iteration number or time $t$ dependence of all trajectories is obtained
by solving the above for $x_{t}$, i.e. 
\begin{equation}
x_{t}=x_{0}\exp _{z}\left[ -\left\vert x_{0}\right\vert ^{1-z}ut\right] .
\label{trajectory}
\end{equation}%
The equivalence of the trajectory positions $x_{t}$ with the size-rank
distribution $N(k)$ is made clear by comparison of Eqs. (\ref{logz1}) and (%
\ref{trajectory}) with Eqs. (\ref{logalpha1}) and (\ref{size-rank1}),
respectively, together with the identifications $t=k$, $u=\mathcal{N}^{-1}$%
, $x_{t}=-N(k)$, $x_{0}=-N_{\max }$ and $z=\alpha $. Also, comparison of the
right-hand side of Eq (\ref{tangent2}) with that of Eq (\ref{cumulative4}%
), taking into account Eq (\ref{cumulative3}), indicates that the analog
of the frequency-rank distribution $f(k^{\prime} )$ is %(minus)
 the quantity%
\begin{eqnarray}
A_{t} &=&\int_{x_{0}}^{x_{t}}\frac{dx_{\tau }}{-\left\vert x_{\tau }\right\vert
^{z}}  \nonumber \\
&=&\ln_{z}\left\vert x_{0}\right\vert - \ln_{z}\left\vert x_{t}\right\vert ,  \label{area1}
\end{eqnarray}%
where $-x_{t}$ plays the role of $k^{\prime }$. In \cite{robledo1} it is pointed out that
the trajectories given by Eq (\ref{trajectory}) have precisely the analytical form for
all trajectories with generic $x_0$ that are generated by the functional composition
renormalization group fixed-point map \cite{schuster1,hu1} at the tangent bifurcation. And
therefore the areas $A_{t}$ in Eq (\ref{area1}) have also the same property.
That is, all trajectories of the fixed-point map for all $t$ initiated at the generic position
$x_0$ obey Eq (\ref{trajectory}). Also Eq (\ref{area1}) enjoys the degree of universality
given by the fixed-point map.

In Fig \ref{Fig3} we illustrate the iterated map properties for the case $z=\alpha=2$
that translate into the equivalent description of the rank distributions $N(k)$ and
$F(k^{\prime})$.    

%FigureIII %<<<<<<<<<<<<
\begin{figure*}[!h]\scriptsize
  \vspace*{.05in}
  \centering
  \includegraphics[width=0.75\textwidth]{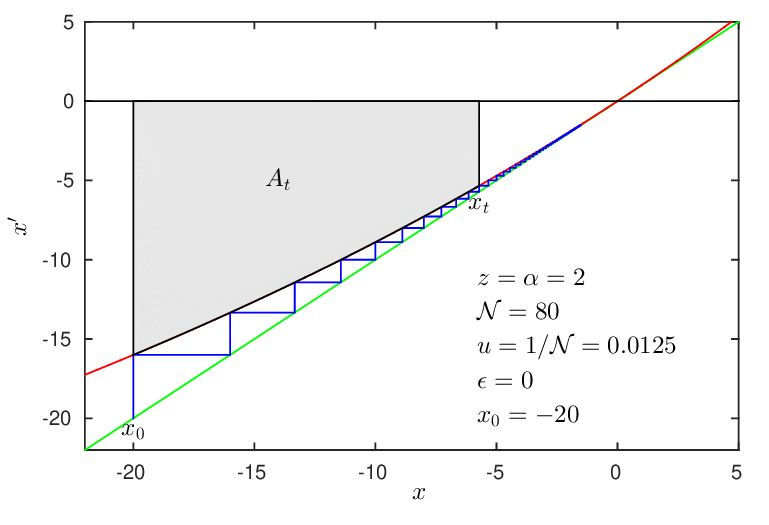}
  \parbox{0.9\textwidth}{\captionFigureIII}
  %\label{Fig3}     %FigureIII(Map_Traj_Area)
\end{figure*}

When $z=1$ we have
\begin{equation}
x_{t}=x_{0}\exp [-ut] .
\label{trajectory2}
\end{equation}
and
\begin{equation}
A_{t} = \ln \left\vert x_{t}\right\vert -\ln\left\vert
x_{0}\right\vert .  \label{area2}
\end{equation}%
The trajectories in Eq (\ref{trajectory2}) are obtained
when a linear map intersects the identity line, i.e.
\begin{equation}
x^{\prime }=f(x)=(1-a)x,  \label{linearmap}
\end{equation}%
and this occurs locally when the tangent map is shifted into a double-secant
map.  

In the limit $z \rightarrow \infty $ the counterpart of Eq (\ref%
{trajectory}) is%
\begin{equation}
x_{t}=\ln \left[ \exp (x_{0})+ut\right] ,  \label{trajectoryinf}
\end{equation}%
as this expression transforms into Eq (\ref{size-rank3}) for $N(k)$ under
the same equivalences $t=k$, $u=\mathcal{N}^{-1}$, $x_{t}=-N(k)$, $%
x_{0}=-N_{\max }$, while that corresponding to Eq (\ref{area1}) is%
\begin{equation}
A_{t}=\exp (x_{0})-\exp (x_{t}).  \label{areainf}
\end{equation}

\section{Rank distributions associated with Benford's first digit law. }

Benford's first digit law \cite{benford1,benford2},
\begin{equation}
\pi(n)=\log(\frac{n+1}{n}),  \label{benford1}
\end{equation}%
where $n$ is the first digit of a decimal base number $N$ and $\log$ denotes the
decimal base logarithmic function, can be readily expressed in terms of the
complementary cumulative distribution Eq (\ref{cumulative}) when $\alpha=1$
and the parent distribution is $P(N)=1/N$. This is 
\begin{equation}
  \pi(n) = \Pi (N,N_{\max }) - \Pi (N+1,N_{\max }) =\log(n+1) - \log(n),  \label{benford2}
\end{equation}%
where $N+1=(n+1).000\cdots$ and $N=n.000\cdots$.

Thus, by considering the cumulative version of Benford's law,
\begin{equation}
\Pi (N,N_{\max})=\log(N_{\max }) - \log(N),  \label{benford3}
\end{equation}%
with $N_{\max}=10$ and $N=n.000\cdots, n=1,2,\ldots,9,$ we have
\begin{equation}
F(k^{\prime }) = \mathcal{N}\log \ (N_{\max}/k^{\prime }), k^{\prime }=1,2,\ldots,N_{\max},
\label{frequency-rankB}
\end{equation}%
and
\begin{equation}
   N(k) = N_{\max} 10^{-k/\mathcal{N}}, k=0,1,\ldots,N_{\max}.  \label{size-rankB}
  % N(k)=N_{\max} \exp (-\mathcal{N}^{-1}k), k=0,1,\ldots,N_{\max}.  \label{size-rankB}
\end{equation}%
In Fig \ref{Fig4} we show these distributions together with numerical data that follows
Benford's law as shown in the figure's inset.

%FigureIV %<<<<<<<<<<<<
\begin{figure*}[!h]\scriptsize
  \vspace*{.05in}
  \centering
  \includegraphics[width=0.75\textwidth]{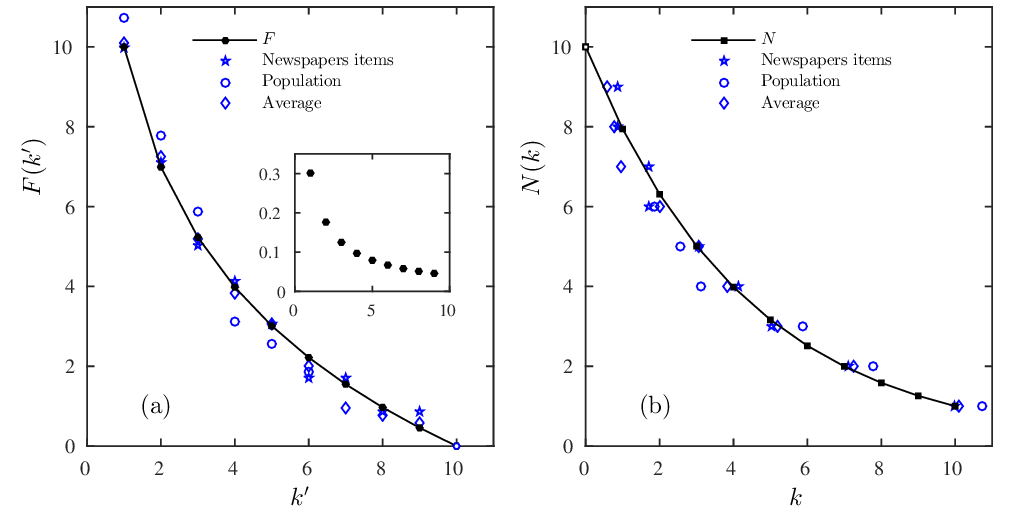}
  \parbox{0.9\textwidth}{\captionFigureIV}
  %\label{Fig4}     %FigureIV(Benford)
\end{figure*}

Benford's law has been generalised to the case $\alpha>1$ \cite{pietronero1}, so
that its associated (complementary) cumulative distribution Eq (\ref{cumulative4})
provides the connection with the rank distributions studied here. In particular the
case $\alpha=2$ corresponds to the classical Zipf's law described by $F(k^{\prime})$
with $k^{\prime}=0,1,2,3,\ldots$, shifted and limited to the values of the first digits
$1,2,3,\ldots,9$, when using decimal base logarithms. 

\section{Rank distributions as expressions of a thermodynamic structure}

As pointed out $N(k)$ is a the functional inverse of $F(k^{\prime})$, that is, the inverse of a
(non-normalized complementary) cumulative distribution in reverse order, a quantile function
\cite{sizerank1,quantile1}. Also $N(k)$ has been interpreted \cite{robledo1, robledo2} as the
total number that the size variable $\it{unit}$ occurs at fixed rank $k$. That is, in thermal system
language, $N(k)$ is equivalent to the degeneracy of a micro state of `energy' $k$, or a
micro-canonical partition function with fixed $k$, where the associated uniform probability is
$p^{(k)}_i=p^{(k)}\equiv1/N(k)$ for all $i=1,\ldots,N(k)$. Thus, we can call $S(k)\equiv\ln N(k)$
an entropy for $\alpha=1$ and define $S(k)\equiv\ln _{\alpha }N(k)$ as a generalized entropy
for $\alpha>1$. Likewise $S_{\max}\equiv\ln N_{\max}$ for $\alpha=1$ and
$S_{\max}\equiv\ln_{\alpha } N_{\max}$ for $\alpha>1$, $N(0)\equiv N_{\max}$. Eq (\ref{logalpha1})
is written now as

\begin{equation}
S_{\max}=S(k)+\mathcal{N}^{-1}k,\label{logalpha2}
\end{equation}%
where, if $S(k)$ is thought of as the entropy for the system with fixed $k$, then 
$S_{\max}(\mathcal{N}^{-1})$ would be a generalized Massieu potential when the variable
$k$ is replaced by the (conjugate) variable $\mathcal{N}^{-1}$ via a Legendre transformation.

Just like thermodynamic quantities are dominant values of statistical-mechanical fluctuating
quantities in a macroscopic system, we think of Eq (\ref{size-rank2}), valid for $\alpha=1$,

\begin{equation}
N_{\max}=N(k)\exp(\mathcal{N}^{-1}k), \label{size-rank5}
\end{equation}
to be the result of the application of the saddle-point approximation for large $\mathcal{N}$ on

\begin{equation}
N_{\max} = \int N(k^{\prime})\exp(\mathcal{N}^{-1}k^{\prime})dk^{\prime}.
\label{size-rank6}
\end{equation}
The consideration of the emergence of dominant rank fluctuations for the general case $\alpha>1$
in the `thermodynamic' limit $k_{\max}=\mathcal{N}\rightarrow \infty$ is less straightforward and
here we do not discuss it further.

We recall \cite{tisza1, callen1} that  the formalism of thermodynamics can be expressed in two equivalent
ways. One of them is to consider as starting point the entropy as the fundamental monotonic function of
the energy (and other basic variables) that characterise the system, while the other alternative is to begin
with the internal energy as the fundamental quantity, a monotonic function of the entropy (and the same
other variables). The expressions obtained from the parent distribution $P(N)$, Eqs. (\ref{logalpha1}) and
(\ref{size-rank1}), correspond to the former choice, while those obtained from the parent distribution $Q(F)$,
Eqs. (\ref{logbeta1}) and (\ref{frequency-rank1b}), relate to the second one. 

% new text added below
In this statistical-mechanical interpretation the rank $k$ plays the role of energy and the entropy is
$S(k)=\ln N(k)$, when $\alpha=1$. In the alternative description $F$ plays the role of energy while
the entropy is $\ln (k^{\prime})$, again when $\alpha=1$. Thus here the quantities representing
entropy and energy are, as customary, functional inverses of each other in accordance with the
usual two equivalent thermodynamic frames \cite{tisza1, callen1}.

As a consequence of the precise analogy between the rank distributions obtained from a parent
distribution and the nonlinear iterated fixed-point map at tangency, we note that the thermodynamic
structure observed above for the rank distributions quantities translates thoroughly into an equivalent
structure for the nonlinear dynamical problem. It is only necessary to recall the identifications $z=\alpha$,
$t=k$, $u=\mathcal{N}^{-1}$, $x_{t}=-N(k)$, $x_{0}=-N_{\max }$ and $A_t=-f(k^{\prime})$ with
$x_{t}=-k^{\prime}$.

%%%%%%%%%%%%%%%%%%%%%%%%%%%%%

\section{Summary and discussion}

We have analyzed the relationship that exists between two types of ranked data, numbers of
occurrences and sizes or magnitudes of items. The technical relationship is well understood for
statistics specialists, frequency-rank data is represented by a (complementary) cumulative
probability distribution while size-rank data is described by its functional inverse, a quantile function
\cite{sizerank1, quantile1}. It is of wider interest, for those studying the many topics of the complex
systems science, where universal patterns are observed in ranked data samples from very different
sources, such as the empirical laws of Zipf and Benford, to understand the physical origin of the
documented behavior We have obtained expressions for size-rank $N(k)$ and frequency-rank
$F(k^{\prime})$ distributions from a stochastic method and or from an equivalent nonlinear
deterministic approach and corroborated that the two functions are inverses of each other.
Their differences are most apparent when the exponent $\alpha$ of the power-law parent
distribution differs from $\alpha=2$, but they coincide and behave as hyperbolic functions
(with deviations for small and large rank) when $\alpha=2$. In this latter case we have a
nonlinear map at tangency with nonzero curvature, the most common case of analytic map
at tangency. This being the case of the classic Zipf law. On the other hand we illustrated the case 
when $\alpha=1$ with the first digit Benford law.

We complemented our description by also considering the option for the parent distribution for the
source of data to be that for the number of occurrences $F$ instead of that for the size $N$.
When these two distributions are assumed to have the power-law forms $P(N)\sim N^{-\alpha }$
and $Q(F)\sim F^{-\beta }$ we obtain parallel (and equivalent) descriptions for the size and
frequency rank distributions with the roles of cumulative distribution and quantile function interchanged
and with the exponents relationship $1-\alpha=(1-\beta)^{-1}$. Further, we advanced a thermodynamic
and statistical-mechanical interpretation to be associated with the properties obtained for the rank
distributions and indicated that the rank $k$  plays the role of energy and $N(k)$ takes the place in a
prototypical thermal system of the number of configurations at fixed energy $k$ with entropy
$S(k)=\ln N(k)$ when $\alpha=1$. The interpretation of the alternative description corresponds to
$F$  playing the role of energy while $\ln (k^{\prime})$ that of entropy when $\alpha=1$. Thus entropy
and energy as functional inverses of each other provide two equivalent thermodynamic formalisms
\cite{tisza1, callen1}.        

The case $\alpha>1$ suggests the use of the generalized entropy expression $S(k)=\ln_{\alpha} (k)$ in
the thermodynamic description, but this poses a question for its corresponding statistical-mechanical
formalism in that the validity of the usual saddle-point approximation requires reconsideration. The outcome
may be one in which fluctuations are not suppressed in the thermodynamic limit, here represented by
$k_{\max }\rightarrow \infty $ and $\mathcal{N}\rightarrow \infty$. The reproduction of the
rank distributions via a nonlinear map at a tangent bifurcation indicates a reason for the
appearance of generalized entropy expression through the drastic contraction of configuration space
from a real number set of possible iterated map trajectories positions to only a finite number in the
limit $k_{\max }\rightarrow \infty $ and $\mathcal{N}\rightarrow \infty$ that corresponds to
$t\rightarrow \infty $ \cite{pnas1, heliyon1}.     

\section {Acknowledgements} AR acknowledges financial support from UNAM PAPIIT
grant IN104417 (Mexican agency).

%%%%%%%%%%%%%%%%%%%%%%%%%%%%%%%%%%%%%%%%%%%%%%%%%%%%%%%%%%%%%%%%%%%%%


\begin{thebibliography}{10}

\bibitem{ranking1}
\newblock To Honor G.K. Zipf
\newblock Glottometrics 3,4,5 Ludenscheid: RAM-Verl., 2002 ISSN 1617-8351.
Available from: \url{http://www.ram-verlag.eu/journals-e-journals/glottometrics/}

\bibitem{newman1}
Newman M.E.J.
\newblock Power laws, Pareto distributions and Zipf's law.
\newblock Contemporary Physics 2005;46(5);323-351.

\bibitem{zipf1}
\newblock Zipf's law. Wikipedia. Available from: \url{https://en.wikipedia.org/wiki/Zipf}.

\bibitem{zipf2}
Zipf G. K.
\newblock Human Behavior and the Principle of Least Effort.
\newblock Cambridge: Addison-Wesley, 1949.

\bibitem{frequencyrank1}
\newblock Frequency-rank distributions. Wikipedia. Available from: \url{https://en.wikipedia.org/wiki/Cumulative_frequency_analysis}.

\bibitem{sizerank1}
\newblock Rank-size distributions. Wikipedia. Available from: \url{https://en.wikipedia.org/wiki/Rank-size_distribution}.

\bibitem{pietronero1}
Pietronero L. Tosatti E. Tosatti V. Vespignani A.
\newblock Explaining the uneven distribution of numbers in nature: the laws of
  {B}enford and {Z}ipf.
\newblock Physica A. 2001;293(1-2):297--304.

\bibitem{robledo1}
Altamirano C. Robledo A.
\newblock Possible thermodynamic structure underlying the laws of {Z}ipf and
  {B}enford.
\newblock Eur\ Phys\ J\ B. 2011;81(3):345--351.

\bibitem{robledo2}
Robledo A.
\newblock Laws of {Z}ipf and {B}enford, intermittency, and critical
  fluctuations.
\newblock Chinese Sci\ Bull. 2011;56(34):3645--3648.

\bibitem{quantile1}
\newblock Quantile function. Wikipedia. Available from: \url{https://en.wikipedia.org/wiki/Quantile_function}

\bibitem{egghe1}
Egghe L. Waltman L.
\newblock Relations between the shape of a size-frequency distribution and the shape of a rank-frequency distribution
\newblock Information Processing and Management 2011;47:238--245.

\bibitem{pnas1}
Yalcin G.C. Robledo A. Gell-Mann M.
\newblock Incidence of $q$ statistics in rank distributions.
\newblock Proc\ Natl\ Acad\ Sci\ USA. 2014;111(39):14082--14087.

\bibitem{schuster1}
Schuster H.G.
\newblock Deterministic Chaos. An Introduction.
\newblock VCH Publishers, Weinheim; 1988.

\bibitem{quakes1}
Southern California Earthquake Data Center. Available from \url{https://www.data.scec.org}

\bibitem{fires1}
Forest Fires Data. Clauset home page Data and Code. Santa Fe Institute.
 Available from: \url{http://tuvalu.santafe.edu/\~aaronc/powerlaws/data.htm}.
  
\bibitem{benford1}
\newblock Benford's law. Wikipedia. Available from: \url{https://en.wikipedia.org/wiki/Benford's law}

\bibitem{benford2}
Benford, F.
\newblock The law of anomalous numbers.
\newblock Proc Am Phil Soc, 1938; 78: 551--572.

\bibitem{heliyon1}
Yalcin G.C. Velarde C. Robledo A.
\newblock Generalized entropies for severely contracted configuration space.
\newblock Heliyon, 2015; 1:e00045. Available from: \url{http://www.heliyon.com/article/e00045/}

\bibitem{hu1}
Hu B. Rudnick J.
\newblock Exact solutions to the Feigenbaum renormalization-
group equations for intermittency.
\newblock Phys Rev Lett, 1982; 48: 1645--1648.

\bibitem{tisza1}
Tisza L.
\newblock Generalized thermodynamics.
\newblock Cambridge University Press, Cambridge; 1993.

\bibitem{callen1}
Callen H.B.
\newblock Thermodynamics and an introduction to thermostatistics.
\newblock Wiley, Cambridge; 1993.

\end{thebibliography}
\end{document}